\documentclass{iau} 
\usepackage{graphicx}

\title[] 
{New insights on the origin of multiple stellar populations in globular clusters}

\author[Jaeyeon Kim \& Young-Wook Lee] 
{Jaeyeon Kim
 \and Young-Wook Lee}

\affiliation{Center for Galaxy Evolution Research \& Department of Astronomy, \\ Yonsei University, Seoul, Korea
 \\ email: {\tt jaeyeonkim93@gmail.com}, {\tt ywlee2@yonsei.ac.kr}}

\pubyear{2017}
\volume{334}  
\jname{Rediscovering our Galaxy}
\editors{C. Chiappini, I. Minchev, E. Starkenburg \& M. Valentini, eds.}
\begin{document}

\maketitle

\begin{abstract}
In order to investigate the origin of multiple stellar populations in the halo and bulge of the Milky Way, we have constructed chemical evolution models for the low-mass proto-Galactic subsystems such as globular clusters (GCs). Unlike previous studies, we assume that supernova blast waves undergo blowout without expelling the pre-enriched gas, while relatively slow winds of massive stars (WMS), together with the winds and ejecta from low and intermediate mass asymptotic-giant-branch stars (AGBs), are all locally retained in these less massive systems. We find that the observed Na-O anti-correlations in metal-poor GCs can be reproduced, when multiple episodes of starbursts are allowed to continue in these subsystems. Specific star formation history (SFH) with decreasing time intervals between the stellar generations, however, is required to obtain this result, which is in good agreement with the parameters obtained from our stellar evolution models for the horizontal-branch. The ``{mass budget problem}" is also much alleviated by our models without \textit{ad-hoc} assumptions on star formation efficiency (SFE) and initial mass function (IMF). We also applied these models to investigate the origin of super-helium-rich red clump stars in the metal-rich bulge as recently suggested by \cite[Lee \etal\ (2015)]{Lee2015}. We find that chemical enrichments by the WMS can naturally reproduce the required helium enhancement ($\Delta{Y}/\Delta{Z}=6$) for the second generation stars. Disruption of proto-GCs in a hierarchical merging paradigm would have provided helium enhanced stars to the bulge field. 
\end{abstract}

\begin{figure}[b]
\begin{center}
 \includegraphics[width=3.8in]{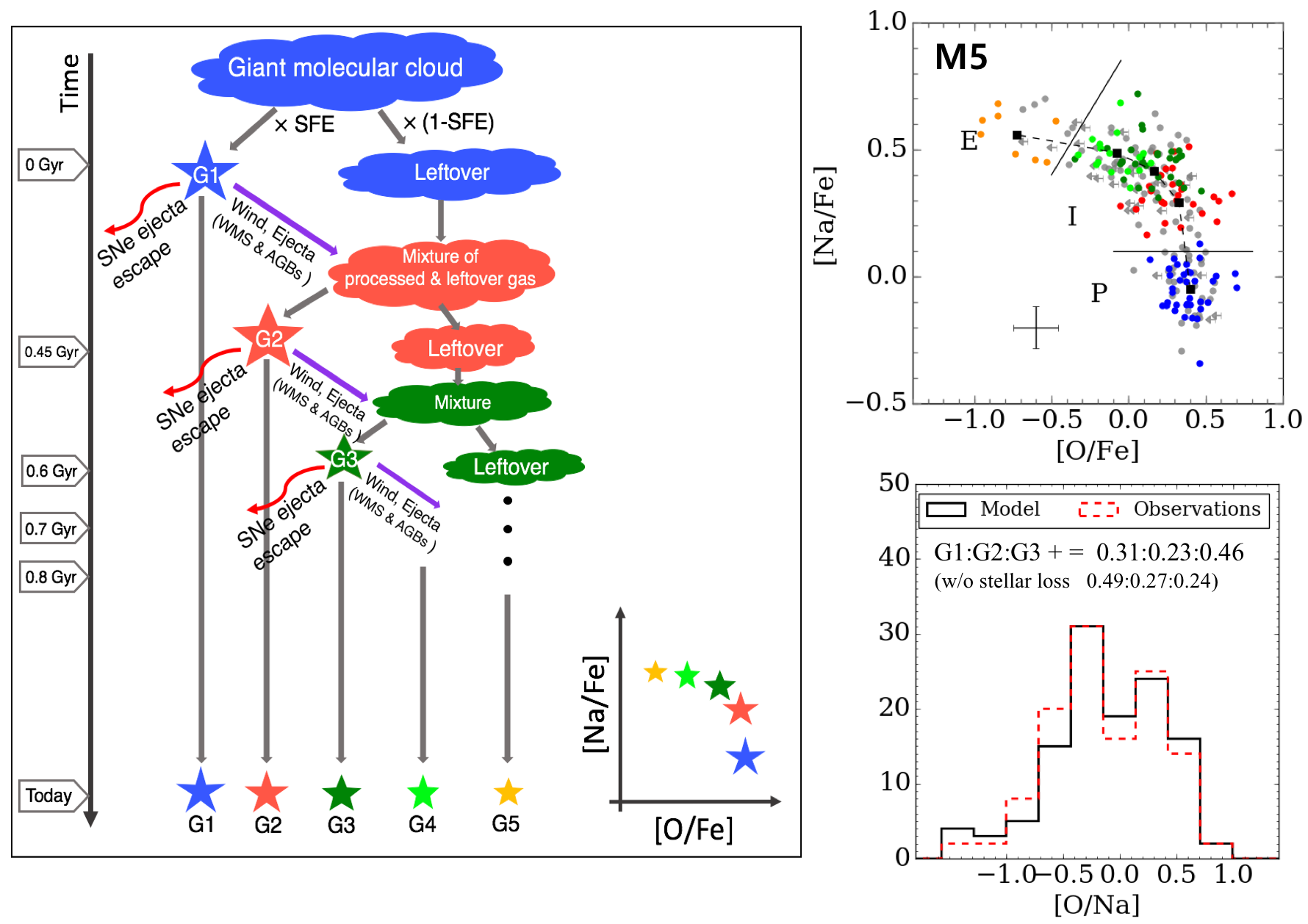} 
 \caption{\textit{Left}: Schematic illustration of our model construction. \textit{Right}: The first, second and later generation stars (G1, G2 and G3\,...) from our chemical evolution models are compared with the observed data from \cite[Carretta \etal\ (2009a,b]{Carretta2009a}; gray scale) for the GC M5.}
   \label{fig1}
\end{center}
\end{figure}

\begin{figure}[b]
\begin{center}
 \includegraphics[width=5.3in]{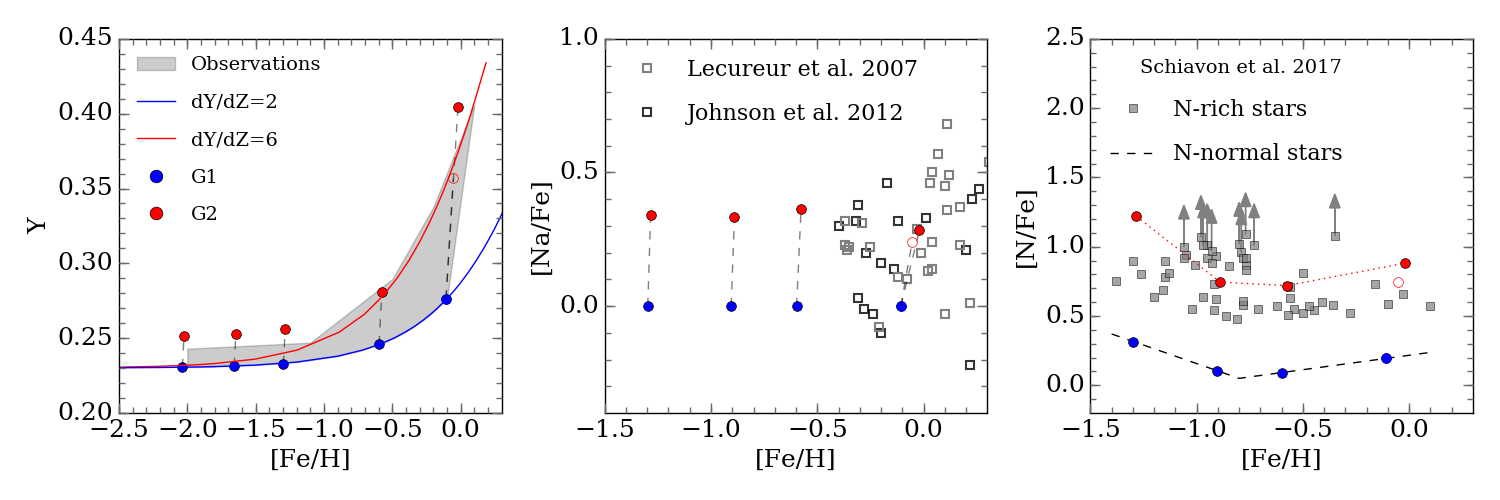} 
 \caption{\textit{Left}: Comparison of He abundances of G1 and G2 from our models with those estimated from stellar evolution models and observations. \textit{middle}: [Na/Fe] abundances of G1 and G2 are superimposed with those of the metal-rich ( [Fe/H] ${>}$ -0.5) bulge stars. \textit{right}: [N/Fe] abundances of G1 and G2 are compared with the N-rich stars observed in the bulge.}
   \label{fig2}
\end{center}
\end{figure}

Our chemical evolution model differs fundamentally from previous approaches (e.g. AGB \&  FRMS scenarios) in that SNe ejecta escapes the system without affecting pre-enriched gas in the proto-GCs (\cite[Tenorio-Tagle \etal\  2015]{Ten2015}; \cite[Silich \& Tenorio-Tagle 2017]{Silich2017}). Furthermore, discrete star forming episodes beyond G2 are allowed to continue to G3 and later generations. Specific SFH was required to best match the observed Na-O anti-correlation of globular clusters (see Fig.\,\ref{fig1} for the case of M5). This is because Oxygen is mostly depleted by WMS and intermediate \& high mass AGBs (5\(M_\odot\) ${\lessapprox}$ M ${\lessapprox}$ 8\(M_\odot\)), whereas Na is mainly enriched by relatively less massive AGB stars (3\(M_\odot\) ${\lessapprox}$ M ${\lessapprox}$ 5\(M_\odot\)). In order to better match the observed [O/Na] histogram, it was required to assume that some of the earlier generation stars were preferentially lost. Nevertheless, the ``\textit{mass budget problem}" is much alleviated by our models without \textit{ad-hoc} assumptions on SFE and IMF. 

Because of the strong metallicity dependence of He yields from WMS (\cite[Maeder 1992]{M92}; \cite[Meynet 2008]{Meynet2008}), our models also predict a large difference in He abundance between G2 and G1 at metal rich regime (see Fig.\,\ref{fig2}), which is required to reproduce the observed double red clump in the Milky Way bulge (\cite[Lee \etal\ 2015]{Lee2015}; \cite[Lee \& Jang 2016]{Lee2016}; \cite[Joo \etal\ 2017]{Joo2017}). Moreover, the observed spreads in [Na/Fe] and [N/Fe] among bulge RGB stars are consistent with our model predictions (Fig.\,\ref{fig2}). These would suggest that proto-globular clusters contributed to the classical bulge formation in the context of hierarchical merging paradigm.

\end{document}